\documentclass[aps,groupedaddress,superscriptaddress,amsmath,amssymb,twocolumn,prx]{revtex4-2}
\usepackage{bm}
\usepackage[dvips]{graphicx}
\usepackage{latexsym}
\usepackage{amsthm}
\usepackage{amsfonts}

\usepackage[colorlinks=true,citecolor=blue,linkcolor=red,urlcolor=blue, linktocpage=true,breaklinks=true,pagebackref=false]{hyperref}

\usepackage{textcomp} 		
\usepackage{hyperref} 		
\usepackage{ifthen} 			
\usepackage{xifthen} 		
\usepackage{color} 			
\usepackage{soul} 			
\usepackage{lineno}
\usepackage[normalem]{ulem}
\usepackage{siunitx}
\usepackage{xcolor} 
\usepackage{float}

\usepackage{braket}

\newcommand{\uproman}[1]{\uppercase\expandafter{\romannumeral#1}}

\newcommand{\addDB}[1]{\textcolor{brown}{#1}}

\begin{document}
	

\title{Frequency comb from a single driven nonlinear nanomechanical mode}
\author{J.\,S. Ochs}
\thanks{Formerly Huber}
\affiliation{Department of Physics, University of Konstanz, 78457 Konstanz, Germany}
\author{D.\,K.\,J. Boneß}
\affiliation{Department of Physics, University of Konstanz, 78457 Konstanz, Germany}
\author{G. Rastelli}
\affiliation{INO-CNR BEC Center and Dipartimento di Fisica, Universit{\`a} di Trento, 38123 Povo, Italy}
\author{M. Seitner}
\affiliation{Department of Physics, University of Konstanz, 78457 Konstanz, Germany}
\author{W. Belzig}
\affiliation{Department of Physics, University of Konstanz, 78457 Konstanz, Germany}
\author{M.\,I. Dykman}
\affiliation{Michigan State University, East Lansing, MI 48824, USA}
\email{dykmanm@msu.edu}
\author{E.\,M. Weig}
\affiliation{Department of Physics, University of Konstanz, 78457 Konstanz, Germany}
\affiliation{Department of Electrical and Computer Engineering, Technical University of Munich, 85748 Garching, Germany}
\affiliation{Munich Center for Quantum Science and Technology (MCQST), 80799 Munich, Germany}
\affiliation{TUM Center for Quantum Engineering (ZQE), 85748 Garching, Germany}
\email{eva.weig@tum.de}

\begin{abstract} 
Phononic frequency combs have been attracting an increasing attention both as a qualitatively new type of nonlinear phenomena in vibrational systems and from the point of view of applications. It is commonly believed that at least two modes must be involved in generating a comb. In this paper we demonstrate that a comb can be generated by a single nanomechanical mode driven by a resonant monochromatic drive. The comb emerges where the drive is still weak, so that the anharmonic part of the mode potential energy remains small. We relate the effect to a negative nonlinear friction induced by the resonant drive, which makes the vibrations at the drive frequency unstable. We directly map the trajectories of the emerging oscillations in the rotating frame and show how these oscillations lead to the frequency comb in the laboratory frame. The results go beyond nanomechanics  and suggest a qualitatively new approach to generating tunable frequency combs in single-mode vibrational systems.
\end{abstract}
	
	\date{\today}
	
	\maketitle
%
%
%
%
\section{Introduction} 
%
%
%
%
%
Since their discovery at the turn of the 21st century~\cite{Cundiff2001,Udem2002}, frequency combs have revolutionized the field of metrology, from unprecedentedly accurate timekeeping to molecule sensing to distance measurements ~\cite{Fortier2019,Picque2019,Chang2022}. 
Frequency combs consist of series of narrow spectral lines \cite{Cundiff2001,Udem2002}. Of central importance for metrology is that the lines are equally spaced. This feature is a consequence of the strong nonlinearity of the radiation sources. 
In lasers such nonlinearity can lead to mode locking which, in turn, leads to generation of frequency combs. It also underlies the electro-optic comb generation.  
In laser-driven microresonators, broad-band optical combs with extremely narrow peaks  emerge as a result of four-wave mixing induced by the Kerr nonlinearity \cite{Del'Haye2007,Kippenberg2011}. A frequency comb associated with a strongly nonlinear parametric excitation has been seen in a superconducting microwave cavity \cite{Erickson2014}.

Frequency combs have also been observed in nanomechanical vibrational systems \cite{Erbe2000,Karabalin2009,Savchenkov2011,Cao2014,Ganesan2017a,Seitner2017,Czaplewski2018,Wei2019,Houri2019,Park2019,Qi2020,Chiout2021,Houri2021,Keskekler2022}. 
Such combs are often called phononic. They cover a broad frequency range and have the advantage of being tunable {\em in situ}.  Their observations have been done by nonlinearly mixing  two drive frequencies, or using nonlinear resonance between different vibrational modes, or using avoided mode crossing. For a single-frequency drive it was substantial that at least two modes are involved, somewhat reminiscent of the multi-mode frequency combs of laser radiation. 
	
The analysis of the frequency combs has led to a general conclusion \cite{Khan2018,Qi2020,Lu2021} that it is necessary to have at least two coupled modes, with at least one of them driven, to generate a comb. This conclusion can be understood by noting that coupled modes can resonantly exchange energy with each other,  an observation which, for nonlinear coupling, goes back to Laplace and Poincar{\`e} on the classical side and to the Fermi resonance, on the quantum side \cite{Arnold1989,Fermi1931}.  The ensuing oscillations are sustained by the external periodic drive. They are slow compared to the mode frequencies and the drive frequency, but can be strongly non-sinusoidal so that their spectrum consists of multiple equidistant lines separated by their frequency. When superposed on the forced vibrations at the drive frequency, this spectrum transforms into a frequency comb.

In this paper we demonstrate that the onset of a frequency comb in a vibrational system does not require mode-mode coupling. We observe a comb using a nanomechanical resonator driven by a single-frequency resonant drive in the regime where only one mode is involved in the dynamics.  The  occurrence of the comb results primarily from the combination of two factors. First, resonant driving can open a relaxation channel that leads to a negative friction force \cite{Dykman2019}. It makes the state of stationary forced vibrations at the drive frequency unstable. As a result, the mode starts precessing about this state. In the frame that rotates at the drive frequency  the precession looks like vibrations. The second  factor is that these vibrations are strongly non-sinusoidal, which leads to a frequency comb, as in the case of coupled modes. As we show, the number of pronounced spectral lines in the comb and the spacing  betwen them can be controlled just by varying the amplitude and frequency of the drive.     

The observation of the comb has been facilitated by our  nanomechanical vibrational mode being  weakly damped. Because the damping is weak, even a comparatively weak driving-induced negative friction force can overcome it, leading to an instability of the stationary forced vibrations. For the same reason, the nonlinearity of the mode comes into play already for a comparatively weak driving provided the driving is resonant. A feature of the nonlinearity is that, in a sense, it is weak: the nonlinear part of the vibration energy is much smaller than the harmonic part. However,  as viewed in the rotating frame, where the harmonic part is largely compensated, 
the nonlinearity can be strong, because it is competing with the weak damping. It is  the strong nonlinearity in the rotating frame that makes the vibrations in this frame strongly non-sinusoidal, leading to multiple lines in the comb spectrum. 

The occurrence of a negative friction force in vibrational systems is
well-known for non-resonant driving \cite{Dykman1978}; it plays an important role in cavity optomechanics \cite{Aspelmeyer2014a}. However, negative friction may also emerge \cite{Dykman2019} for resonant driving~\footnote{Preliminary experimental data on the onset of a frequency comb in a nanomechanical resonator were presented in the thesis \url{https:} and were shown by E.~M.~Weig at the conference on Frontiers of Nanomechanical Systems 2019, \url{https://fns2019.caltech.edu}}, as also suggests the experiment \cite{Bousse2020}. The resonantly induced friction force (RIFF) is nonlinear in the mode coordinate. Therefore, it comes into play only once the vibration amplitude becomes sufficiently large. We observe its effect only on the large-amplitude branch of the response curve.

%
%
\section{Setup and Characterization}
\begin{figure}
	\includegraphics[width=0.95\linewidth]{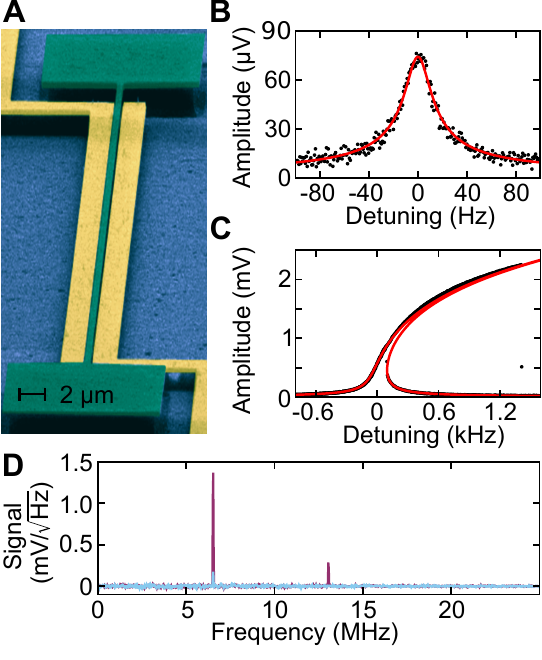}
	\caption{Nanomechanical string resonator in the single-mode regime (a) Scanning electron micrograph of the doubly clamped silicon nitride string resonator (green) and two control electrodes (yellow). (b) Linear response for a drive power of $-56$\,dBm along with a Lorentzian fit (solid red line). A constant noise background is subtracted from the data ($1.5 \cdot 10^{-6}$\,V). (c) Amplitude of forced vibrations for a moderately strong drive power of $-24$\,dBm as a function of drive frequency along with a fit (solid red line, see SM). (d) Vibration spectra for a resonant drive applied at the OOP mode in the linear regime ($-46$\,dBm, light blue), and for a stronger drive ($5$\,dBm,  purple) that exhibits an overtone at twice the drive frequency. Even under strong driving, no other modes are excited. A noise background has been subtracted from the data.
	}
	\label{fig:fig1}
\end{figure}

The investigated nanomechanical resonator is a freely-suspended string resonator, similar to the one depicted in Fig.\,\ref{fig:fig1}\,(a). It is fabricated from pre-stressed silicon nitride  on a fused silica substrate, facilitating ultra-high quality factors $\gtrsim 10^5$ at room temperature. The string has a length of $55\,\si{\micro \meter}$, a width of $270\,\si{\nano \meter}$, and a thickness of $100\,\si{\nano \meter}$.
Integrated dielectric transduction combined with microwave cavity-enhanced heterodyne detection, described in Refs. \cite{Faust2012,Rieger2012,Unterreithmeier2009}, is implemented via the two gold electrodes also apparent in Fig.\,\ref{fig:fig1}\,(a) and allows actuating and detecting the  motion of the resonator. All measurements are performed at a constant DC voltage of $5\,\si{\volt}$, under vacuum at a pressure of $\leq10^{-4}\,\si{\milli \bar}$ and at room temperature of $293\,\si{\kelvin}$.

%

The fundamental flexural out-of-plane (OOP) mode of the resonator is characterized in the linear as well as in the nonlinear regime. This is done by driving the mode with a single-tone drive $F\cos(2\pi f_d t)$ applied on or near resonance.  From the linear response measurement (black dots) and a Lorentzian fit (solid red line), shown in Fig.\,\ref{fig:fig1}\,(b), one obtains the mode eigenfrequency $f_{0}=\omega_0/2\pi\approx 6.528$\,MHz, the linewidth $2\Gamma/2\pi\approx 21$\,Hz,  and with that the Q-factor of  $Q\approx 310,000$. 

The  nonlinear response of the resonator as a function of the detuning $f_d - f_0$ is shown in Fig.\,\ref{fig:fig1}\,(c). In the presented parameter range the forced vibrations were nearly sinusoidal. The stationary value of the mode coordinate is $q(t) \approx A \cos (2\pi f_d t + \phi)$, where $A$ and $\phi$ are the vibration amplitude and phase.

In nanomechanics, resonant nonlinear response is most frequently described by the Duffing (sometimes also called Kerr) model, in which the nonlinear part of the potential of the mode $U(q)$  has the form $M\gamma q^4/4$ \cite{Bachtold2022}, where $M$ is the effective mass of the mode and $\gamma$ is the Duffing parameter. The response in Fig.\,\ref{fig:fig1}\,(c) cannot be fit to the standard Duffing curve. The deviation is due to the resonator lacking inversion symmetry, for example due to the term $\propto q^3$ in $U(q)$ \cite{Landau2004a}. For the studied mode it leads to a significant reduction  of the positive ``bare'' value of $\gamma$, i.e., $\gamma \to \gamma_{\mathrm{eff}}$, in the regime of comparatively small vibration amplitudes \cite{Ochs2021}.  

In the experiment the mode amplitude and the amplitude of the driving force are measured in volts. In these units, the renormalized  value of the Duffing parameter at comparatively small vibration amplitudes $\lesssim 1$~mV is $\gamma^\mathrm{(V)}_\mathrm{eff}/(2\pi)^2 \approx 2.57\cdot 10^{15}\mathrm{V^{-2}s^{-2}}$. The full theoretical  analysis, including the full response curve in Fig.~\ref{fig:fig1}\,(c), is given in Sec.~\ref{sec:Interpretation}, see also Supplemental Material (SM).

The DC voltage is chosen in such a way as to operate the system in the single-mode regime. No other modes are excited where the OOP mode was driven on or close to resonance. Figure\,\ref{fig:fig1}\,(d) displays the vibration spectrum for a drive applied at the eigenfrequency of the OOP mode. In the linear response regime, only forced vibrations at the drive frequency are observed (light blue line). For a stronger drive, we observe an overtone at twice the drive frequency (purple line), as expected for a resonator with broken inversion symmetry. We do not see signals at other frequencies. The full mode spectrum of the device is discussed in the Supplementary Material (SM). 
%
%
%
\section{Experimental Observations}
\label{sec:experiment}
%
%
%
\begin{figure}
	\centering
	\includegraphics[width=0.95\linewidth]{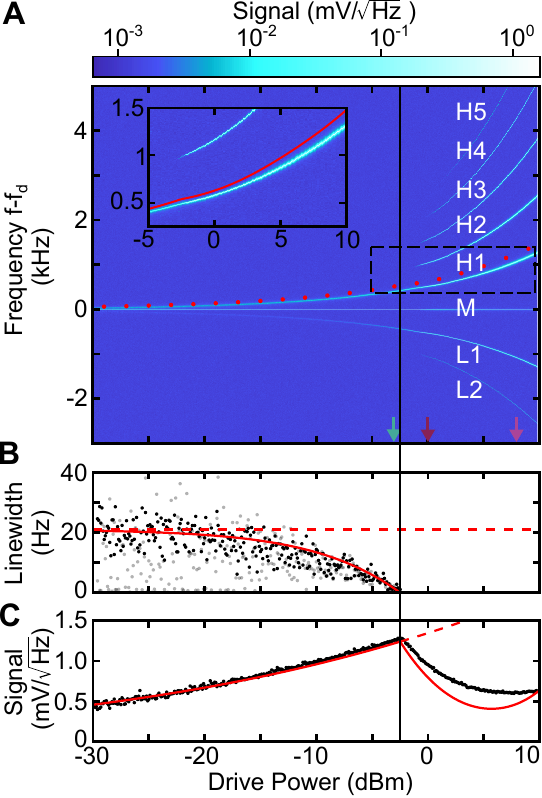}
	\caption{ Frequency comb induced by resonant drive (a) Power spectra of the mode measured for increasing drive power at $f_d = f_0$. The transition from the regime of two thermal-noise-induced satellite peaks to the frequency comb consisting of a series of equidistantly-spaced, multiple satellite peaks is clearly seen. It occurs at $\mathcal{P}_\mathrm{th}=-2.5$\,dBm (black vertical line). The red dots show the theoretical model for H$_1$. The inset shows the magnified region inside the dashed box. Arrows indicate drive power of the linecuts depicted in Fig.~\ref{fig:fig3}(a). (b) Extracted linewidth of the two thermal noise-induced satellite peaks H$_1$ (black) and L$_1$ (gray) as a function of the drive power. Dashed line represents the prediction based on the theory that disregards driving induced friction, whereas the solid line includes it, see Eqs.~(\ref{eq:RIFF}) and (\ref{eq:drive_induced_decay}). (c) Amplitude of forced vibrations at $f_d=f_0$ as a function of the drive power. Dashed line represents the prediction based on the theory that disregards driving induced friction, whereas the solid line includes it, see Eqs.~(\ref{eq:RIFF}) and (\ref{eq:drive_induced_decay}).
	}
	\label{fig:fig2}
\end{figure}
%
%
%
\begin{figure*}
	\centering
	\includegraphics[width=0.95\linewidth]{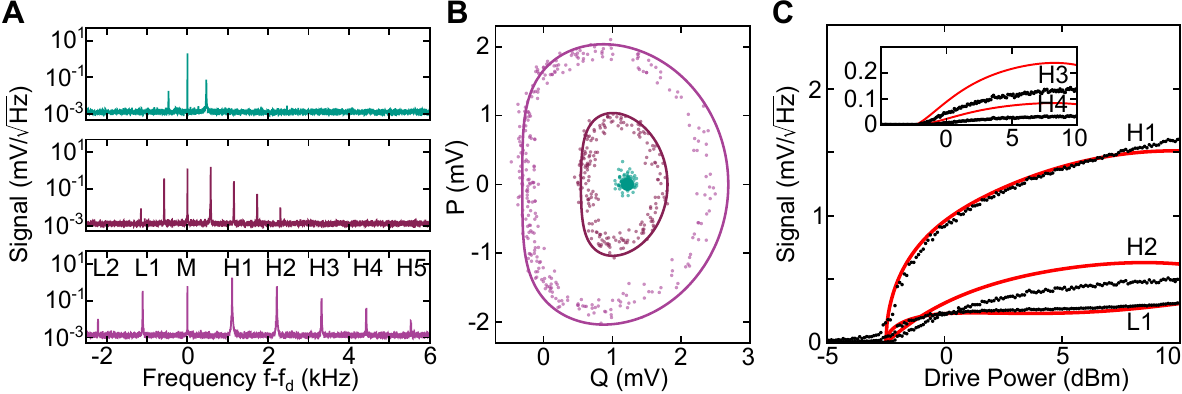}
	\caption{Spectral vs. homodyne measurement. (a) Frequency combs at drive power $-3$\,dBm (top), $-1$\,dBm (center), and $8$\,dBm (bottom). Data is taken from linecuts of Fig.\,\ref{fig:fig2}\,(a) indicated by small arrows. 
	(b) Trajectories in the rotating frame for the same drive powers as in (a). Data has been rescaled as described in the SM. Solid lines depict theoretical model.
	(c) Amplitude of the satellites L$_1$, H$_1$, H$_2$ as well as H$_3$ and H$_4$ (magnified in inset) from Fig.\,\ref{fig:fig2}\,(a) (black dots), compared to the Fourier components obtained from the theoretical model (red lines).
	}
	\label{fig:fig3}
\end{figure*}

\subsection{Driving on resonance}
\label{subsec:resonant_drive}

In Fig.~\ref{fig:fig2}\,(a) we show the power spectra of the OOP mode driven sharp on resonance, $f_d=f_0$, for the drive power in the range from $-30$\,dBm to $+10$\,dBm. For drive powers below $-15$\,dBm, besides the main-tone (M) peak at the drive frequency $f_d$, we observe two thermal-noise-induced satellite peaks, which appear in the spectrum symmetrically to the left (L$_{1}$) and right (H$_{1}$) of M. Such satellites, as well as the spectral evidence of the squeezing encoded in their unequal brightness, is discussed in Ref.~\cite{Huber2020}.

As the drive power approaches  the threshold value $\mathcal{P}_\mathrm{th} \approx -2.5$\,dBm (black vertical line), the noise-induced satellites (H$_1$, L$_1$)  evolve into much narrower peaks. For increasing power, we resolve additional, equally spaced satellite peaks forming a frequency comb.  We label them as (H$_2$, L$_2$), (H$_3$, L$_3$), etc., with H referring to the higher- and L to the lower-frequency satellites. Note that the lower-frequency satellites are less intense than the higher-frequency ones.

The linewidth of the first higher- and lower-frequency peaks H$_1$ and L$_1$ is explored in more detail in Fig.\,\ref{fig:fig2}\,(b). The linewidth is found from a Lorentzian fit. It remains at a constant value of  $\approx 21$\,$\si{\Hz}$ up to a drive power of $-15$\,dBm. This value corresponds to the linear damping rate of the OOP mode.
%
For larger drive powers, the linewidth gradually decreases until at  $\mathcal{P}_\mathrm{th}$ it reaches the resolution limit  $1$\,$\si{\Hz}$ of our measurement device. Respectively, in Fig.~\ref{fig:fig2}\,(b) we do not show the linewidth for $\mathcal{P}>\mathcal{P}_\mathrm{th}$. The linewidths of higher-order satellites, which appear for $\mathcal{P}>\mathcal{P}_\mathrm{th}$, also could not be resolved.

Figure\,\ref{fig:fig2}\,(c) extracts the amplitude of the main tone (M), i.e., the amplitude of the forced vibrations at $f_d=f_0$, as a function of the drive power. The signal increases with the drive power until $\mathcal{P}_\mathrm{th}$. 
At the threshold power, the amplitude of the forced vibrations exhibits a kink and abruptly starts to decrease.  

Strikingly, the succinct features of Fig.\,\ref{fig:fig2}\,(b) and (c), the linewidth reaching the resolution limit and the kink in the amplitude of the forced vibrations, coincide with the emergence of the frequency comb at $\mathcal{P}_{\text{th}}$ (black vertical line).

Figure\,\ref{fig:fig3}\,(a) presents linecuts of Fig.\,\ref{fig:fig2}\,(a) at the drive powers indicated in Fig.~\ref{fig:fig2}\,(a) by small arrows. Beyond $\mathcal{P}_\mathrm{th}$, the intensity of the satellite peaks forming the frequency comb strongly increases with the increasing drive strength. It is noteworthy to observe that some satellites even exceed the intensity of the main tone (M). 

In addition to the spectral measurements discussed so far, we directly record the trajectories of the system in the rotating frame using a homodyne measurement. A fast lock-in amplifier is employed to sample both the in-phase and quadrature signal of the driven resonator over time. In order to capture the frequency comb, the measurement is performed with a large bandwidth of $10\,\si{\kilo \Hz}$. Figure\,\ref{fig:fig3}\,(b) plots the obtained trajectories in the space of the in-phase ($Q$) and quadrature ($P$) components, i.e., in the phase space of the rotating frame. For a drive power of $-3$\,dBm, corresponding to the top panel in Fig.\,\ref{fig:fig3}\,(a), the system is still in a stable state of forced vibrations, such that the trajectory mostly stays within a thermal-noise-broadened ellipse centered at the value of $Q$ and $P$ in the stable state. For a larger drive power of $-1$\,dBm, beyond the instability threshold, the system is clearly on a limit cycle. It represents the self-sustained oscillations of the resonator in the rotating frame. The shape of the trajectory is profoundly non-elliptical, which means that the vibrations of $Q(t)$ and $P(t)$ are non-sinusoidal. This is consistent with the observation of multiple satellites in the power spectrum in the middle panel of Fig.\,\ref{fig:fig3}\,(a). For a still larger drive power of $8$\,dBm, a larger and even more non-elliptical limit cycle is observed, which is in line with the bottom panel of Fig.\,\ref{fig:fig3}\,(a).

Interestingly, the trajectories are practically symmetric with respect to the axis $P=0$. As we explain in Sec.~\ref{sec:Interpretation}, this is a consequence of a very small decay rate of the mode. 
The amplitude of the first lower and the first four higher satellites L$_1$, and H$_1$ to H$_4$ is plotted in Fig.\,\ref{fig:fig3}\,(c) as a function of drive power.

\subsection{Frequency comb as a function of the drive detuning}
\label{subsec:detuning}

\begin{figure}
	\centering
	\includegraphics[width=0.95\linewidth]{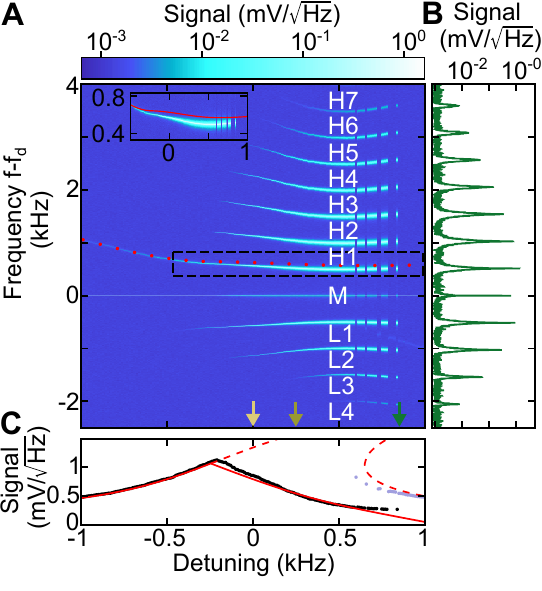}
	\caption{Frequency comb for detuned drive. (a) Power spectra measured for increasing detuning $f_d-f_0$ at a fixed drive power of $0$\,dBm. The dotted line represents the theoretical model for H$_1$. The inset magnifies the region indicated by the dashed box. Arrows indicate drive power of the trajectories depicted in Fig.~\ref{fig:fig5}(a). (b) Frequency comb in linecut of (a) taken at a detuning of $840$\,Hz (green arrow). (c) Amplitude of forced vibrations at $f_d$ as a function of detuning extracted from (a). Black datapoints refer to the large-amplitude solution, whereas light purple indicates the small-amplitude solution for which there is no instability. Dashed line represents the prediction based on the theory that disregards driving induced friction, whereas the solid line includes it, see Eqs.~(\ref{eq:RIFF}) and (\ref{eq:drive_induced_decay}). 
	}
	\label{fig:fig4}
\end{figure}

\begin{figure}
	\centering
	\includegraphics[width=0.95\linewidth]{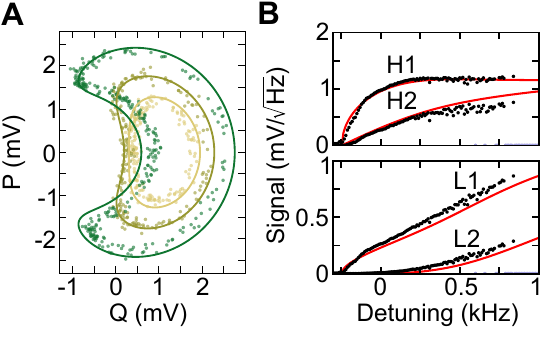}
	\caption{Nonsinusoidal trajectories for detuned drive. (a) Trajectories in the rotating frame for a detuning of $0$\,Hz (light green), $250$\,Hz (medium green), and $850$\,Hz (dark green) at a drive power of $0$\,dBm.  Data has been rescaled as described in the SM.  
		(b) Amplitude of the satellites H$_1$, H$_2$ (top) as well as L$_1$, L$_2$ (bottom) as a function of detuning from Fig.\,\ref{fig:fig4}\,(a) (black dots), compared to the Fourier components obtained from the theory (red lines).
	}
	\label{fig:fig5}
\end{figure}

Along with the onset of the frequency comb with the increasing drive amplitude, we have studied the emergence of a comb with the varying drive frequency $f_d-f_0$. The results for a fixed drive power of $0$\,dBm are shown in Figs.\,\ref{fig:fig4} and \ref{fig:fig5}. For this drive power, and at finite detuning $f_d-f_0>0$, the spectra display an even more elaborate frequency comb than in Fig.~\ref{fig:fig3}. 

For each measurement, the resonator is initialized in the large-amplitude state by sweeping up the drive frequency
from $30$\,kHz below $f_0$ to the desired drive frequency $f_d$ prior to recording the power spectrum. For a large negative detuning the system remains stable for the selected drive power. The instability occurs at a detuning of -$210$\,Hz, and for increasing detuning an increasingly multiple-line frequency comb rapidly evolves. The number of visible lines is significantly larger than for the case of a resonant drive, as also apparent from the linecut in Fig.\,\ref{fig:fig4}\,(b). 

The comb only exists in the detuning range limited by the bifurcational value at which the large-amplitude branch of the response curve disappears as a whole, cf. Fig.~\ref{fig:fig1}~(c). On the small-amplitude branch, the mode does not display a frequency comb. 
The rate of switching to the small-amplitude branch increases exponentially as the system approaches the bifurcation point \cite{Bachtold2022}, and so does the probability of switching during the measurement. The randomness of the switching is manifested in the apparent gaps in the comb and in the jumps in the vibration amplitude of Fig.\,\ref{fig:fig4}\,(a) and (c), respectively. Both signatures result from such switching. The power spectra have been obtained with a $10$\,Hz increment. Therefore, for example, the two gaps and the respective jumps at $600$\,Hz and $660$\,Hz correspond to single random switching events. 
Counterintuitively, the amplitude of the vibrations at the drive frequency sharply increases when the system switches to the small-amplitude branch, as apparent from Fig.\,\ref{fig:fig4}\,(c). The forced vibrations at the drive frequency remain stable on this branch, there is no frequency comb.

Figure\,\ref{fig:fig5}\,(a) depicts three trajectories measured for a detuned drive (color-coded detuning marked by arrows in Fig.\,\ref{fig:fig4}\,(a)), all at the same drive power. For increasing detuning the size and asymmetry of the observed limit cycles clearly increase, giving rise to a horse-shoe like  trajectory. The presented trajectories are significantly more asymmetric than those for a resonant drive, in agreement with the higher number of frequency comb lines observed. 
The amplitude of the first two lower and higher satellite peaks is extracted as a function of the detuning in  Fig.\,\ref{fig:fig5}\,(b).  	
%
%

%
%
%
%
%
%
%
%
%

%
%
\section{Interpretation}
\label{sec:Interpretation}
%
%
%
A remarkable feature of the studied nanomechanical mode is that it is weakly damped not only in the laboratory frame, $\Gamma\ll \omega_0 = 2\pi f_0$, but also in the rotating frame. To understand the experimental observations it is necessary, first, to understand the dynamics of the mode in the absence of dissipation. Understanding the dissipation mechanisms is the next step.


\subsection{The Hamiltonian dynamics of the driven mode}
\label{subsec:Hamiltonian}

We start with the Hamiltonian dynamics in the laboratory frame  and then proceed to the Hamiltonian dynamics in the rotating frame. 
In the studied regime the anharmonic  part of the potential energy of the mode $U(q)$ is small compared to the harmonic part $(p^2/2M) + (M\omega_0^2 q^2/2)$, where $p$ is the momentum in the laboratory frame. The major effect of the mode nonlinearity is the dependence of the vibration frequency on the mode energy $E$, or equivalently, on the action variable $I$, with the frequency being $\omega(I) = dE/dI$ \cite{Landau2004a}. 
In the Duffing model $\omega(I) \approx \omega_0 + \alpha_1 I$ for small $I$, with $\alpha_1 = 3\gamma /4M\omega_0^2$ [$\omega_0\equiv \omega(0)$]. More generally, $\alpha_1$ is proportional to the effective Duffing parameter $\gamma_\mathrm{eff}$ rather than $\gamma$. Thus, $\alpha_1$ is directly accessible from the experiment in the regime of comparatively small vibration amplitudes even if the Duffing model does not hold for increasing amplitudes.

The peculiarity of the studied mode is the significant reduction of $\gamma_\mathrm{eff}$ compared to $\gamma$. It means that $\alpha_1$ is small. Therefore, to describe the response for larger amplitudes it is necessary to keep a higher-order term in the expansion of the frequency in $I$, i.e., to set  $\omega(I) \approx \omega_0 + \alpha_1 I + \alpha_2 I^2$. This corresponds to the Hamiltonian of the mode in the laboratory frame of the form
\begin{align}
\label{eq:H_0_lab_frame}
H_0=\omega_0I + \frac{1}{2} \alpha_1 I^2 + \frac{1}{3}\alpha_2 I^3.
\end{align}
This form is general. It incorporates not only the nonlinearity of the isolated mode, but also the renormalization of the mode parameters due to a nonresonant coupling to other modes of the nanoresonator, including phonons \cite{Bachtold2022}. 
The parameter $\alpha_2$ has contributions quadratic in $\gamma$ and also contributions from the terms $\propto q^3$, $q^5$, $q^6$ in the potential $U(q)$, see SM. However, all these terms are renormalized. The parameters $\alpha_1$ and $\alpha_2$ are the only relevant parameters of the conservative dynamics in the whole range of moderately large vibration amplitudes. 

The effect of the resonant driving force is described by incorporating into the mode Hamiltonian the term $H_F=-qF\cos \omega_dt$ ($\omega_d = 2\pi f_d$). This effect becomes strong already for comparatively weak force amplitude $F$, as $F$ is ``competing'' with the small frequency detuning $|\omega_d - \omega_0|\ll \omega_d$. The mode dynamics can be analyzed using the conventional method of averaging  \cite{Arnold1989}, which in this case is the averaging over the drive period $2\pi/\omega_d$. A significant simplification comes from the fact that, in the considered amplitude range, the mode coordinate in $H_F$ can be approximated as $q\approx (2I/M\omega_d)^{1/2}\cos\theta$, where $\theta$ is the vibration phase. 

The averaged Hamiltonian describes the dynamics of the driven mode in the rotating frame. The mode coordinate $Q$ and momentum $P$ in this frame correspond to the in-phase and quadrature components of the vibrations,
\begin{align}
\label{eq:variable_transformation}
Q + iP = [q + i(p/M\omega_d)]\exp(i\omega_d t).
\end{align}
In these variables the Hamiltonian of the driven mode $g(Q,P)$ reads (see SM)
\begin{align}
\label{eq:H_rotating_frame}
&g(Q,P)=(M\omega_d)^{-1}(H_0-\omega_d I - QF/2),\nonumber\\
&I= M\omega_d (Q^2 + P^2)/2.
\end{align}
The dynamical variables $Q,P$ satisfy the standard Hamiltonian equations $\dot Q = \partial_P g, \dot P=-\partial_Q g$. In quantum terms, the value of $g$ on a Hamiltonian trajectory is the quasienergy, or the Floquet eigenvalue of the driven mode 
\cite{Shirley1965,Zel'dovich1967a,Ritus1967,Sambe1973}. 
Also, the employed method of averaging goes beyond the conventional rotating wave approximation.

We emphasize that, even though the dynamics in the laboratory frame is weakly nonlinear, the dynamics in the rotating frame is strongly nonlinear.  The vibrations $Q(t),P(t)$ with a given $g$ can be strongly nonsinusoidal. The typical frequencies of these vibrations $\nu(g)$ turn out to be much higher than the decay rate of the mode, that is, the mode is underdamped not only in the laboratory frame, but also in the rotating frame. It is this property that determines the shape of the measured trajectories in Figs.~\ref{fig:fig3}~(b) and \ref{fig:fig5}~(a), which are essentially the Hamiltonian trajectories $g(Q,P)=$~const. This means that the measured trajectories directly depict constant-quasienergy contours.  Because $g(Q,P) = g(Q,-P)$, the Hamiltonian trajectories are symmetric with respect to the axis $P=0$. The symmetry of the measured trajectories is a signature of the weak dissipation in the experiment.

For a given $g(Q,P)=g$, the nonsinusoidal trajectories in the rotating frame have multiple equally spaced Fourier components separated by  $\nu(g)$. As seen from Eq.~(\ref{eq:variable_transformation}), these components  modulate the vibrations of the mode at frequency $\omega_d$ in the laboratory frame. Therefore, they are seen in the spectrum of the mode as equidistant lines separated by $\nu(g)$. This underlies the frequency comb observed in the experiment.

\subsection{Instability mechanism}
\label{subsec:instability}

The observation of the instability of forced vibrations with the increasing vibration amplitude and the ``soft'' onset of the frequency comb suggests that the effective friction force turns to zero at the instability threshold \cite{vanderPol1926}. This implies the existence of a negative nonlinear friction force. Such force must be increasing with the increasing vibration amplitude so that at the threshold it compensates the conventional friction force  $-2M\Gamma\dot q$. There are no a priori reasons to expect that, if this force is retarded, the retardation time will be so long as to be comparable with the dynamical times in the rotating frame. Therefore, given that we are interested in the dynamics in the rotating frame, in the phenomenological description retardation can be disregarded. 

A simple phenomenological form of the relevant friction force is RIFF \cite{Dykman2019},
\begin{equation}
\label{eq:RIFF}
F_\mathrm{RIFF} = -\eta_\mathrm{RIFF}F \cos(\omega_d t) \dot{q}(t) q(t).
\end{equation}
The force $F_\mathrm{RIFF}$  has the proper time and spatial symmetry, and the nonlinearity is of the lowest order (quadratic) in the vibration amplitude. A microscopic model of this force was related \cite{Dykman2019} to the fact that the work by the force averaged over the period $2\pi/\omega_d$, $[ F \cos(\omega_d t) \dot{q}(t)]_\mathrm{av}$, leads to heating of a nanoresonator. The associated thermal expansion can reduce tension in the nanostructure and thus the mode eigenfrequency. One can easily infer from Fig.~\ref{fig:fig1}~(c) that the decrease of the eigenfrequency with the increasing vibration amplitude on the upper branch of the response curve can lead to an instability. However, our estimates show that the heating is too weak for the nanoresonator studied in the experiment. Therefore\addDB{,} we consider $\eta_\mathrm{RIFF}$ as an adjustable parameter. Replacing  in Eq.~(\ref{eq:RIFF})  $F \cos(\omega_d t) \dot{q}(t)\to [F \cos(\omega_d t) \dot{q}(t)]_\mathrm{av}$  weakly affects the results, and since such a replacement is physically appealing, we use it.

Besides the RIFF, the resonant drive can open another relaxation channel,  similar to the driving-induced relaxation for a nonresonant drive in cavity optomechanics \cite{Dykman1978,Aspelmeyer2014a}. It can be understood by thinking that the drive modulates the coupling of the mode to a thermal bath, with the interaction Hamiltonian of the form $H_i=F\cos(\omega_dt)q h_\mathrm{b}$, where $h_\mathrm{b}$ depends on the dynamical variables of the bath. The modulation gives rise to the driving-induced decay processes, which change the mode decay rate (see SM),
\begin{align}
\label{eq:drive_induced_decay}
&\Gamma \to \Gamma +F^2\Gamma_d,\nonumber\\
&\Gamma_d = \frac{1}{8\hbar M\omega_0}\,\mathrm{Re}\,\int_0^\infty dt e^{2i\omega_d t}\langle [h_b(t),h_b(0)]\rangle\addDB{.}
\end{align}

The dependence of $F_\mathrm{RIFF}$ on the vibration amplitude plays a dual role. On the one hand, the increase of $F_\mathrm{RIFF}$ with the increasing amplitude leads to the very instability of the stationary state of forced vibrations at $\mathcal{P}_\mathrm{th}$. On the other hand, once the system starts vibrating in the rotating frame, its mean amplitude in the laboratory frame, which is $\propto (Q^2+P^2)^{1/2}$,  decreases. This  is clearly seen in Fig.~\ref{fig:fig2}~(c). Then $F_\mathrm{RIFF}$ decreases with the increasing amplitude of vibrations in the rotating frame. As a result, a stable limit cycle forms in the rotating frame. In the weak-damping limit, it corresponds to vibrations with the value of the quasienergy $g$ such that the dissipative terms including $F_\mathrm{RIFF}$, averaged over the orbit $g(Q,P)=g$, exactly compensate each other. With the increasing drive power this $g$ increases, leading to the evolution of the trajectories in Fig.~\ref{fig:fig3}~(b). We note that vibrations in the rotating frame can be also generated using a feedback loop, as demonstrated by \textcite{Houri2021}.

The above nonlinear theory has 4 parameters, $\alpha_{1,2}$, $\eta_\mathrm{RIFF}$, and $\Gamma_d$. The parameters $\alpha_{1,2}$ are found directly from the measurements of the response curve in Fig.~\ref{fig:fig1}~(c), see SM. The parameters $\eta_\mathrm{RIFF}$ and $\Gamma_d$  can be chosen so as to describe the amplitude of the vibrations at the drive frequency as a function of the drive power shown in Fig.~\ref{fig:fig2}~(c). The value of the drive power at the threshold of the instability $\mathcal{P}_\mathrm{th}$  is particularly sensitive to $\eta_\mathrm{RIFF}$. With these parameters we describe the multitude of the observations, including not only the response, but also the shape of the trajectories and its dependence on the drive power and frequency, Figs.~\ref{fig:fig3}~(b) and \ref{fig:fig5}~(b), as well as the positions and intensities of the comb lines (see SM).

In order to highlight the connection between the number of satellites in the frequency comb and the asymmetry of the trajectory of the limit cycle, Fig.\,\ref{fig:fig3}\,(c) and Fig.\,\ref{fig:fig5}\,(b) plot the amplitude of the Fourier components (red lines) on top of the amplitude of the measured comb lines (black dots). We find good agreement between the two.


\section{Summary and outlook}

We experimentally  demonstrate that a resonantly driven single nanomechanical mode can display self-sustained vibrations in the rotating frame. No coupling to other modes is required. The vibrations  are strongly nonlinear, even though for the studied drive power the anharmonic part of the mode potential (in the laboratory frame) remains much smaller than the harmonic part. The trajectories of the observed limit cycles  in the rotating frame represent constant-quasienergy contours and thus allow one to directly map out its shape as a function of the applied drive power. The Fourier components of the vibrations are manifested in the spectrum of the mode in the laboratory frame as extremely sharp equidistant peaks that form a frequency comb. 

We find that the number of visible lines in the comb and the line spacing sensitively depend on the power and frequency of the close-to-resonance drive. This  suggests a straightforward way of controlling the corresponding parameters, which is important for numerous applications of the phononic frequency combs. Other advantageous features of the system include the low power required to generate the comb and the very frequency range where the comb emerges.

A good agreement of the theory with all experimental observations provides an evidence of a qualitatively new mechanism of dissipation of driven vibrational systems, the resonantly induced friction force. Developing a microscopic theory of this force is a challenging problem that will be addressed in the future work.  \\


%
%
%
\section{Acknowledgements}
We are grateful to S.\,W.\,Shaw for a discussion. J.\,S.\,O., D.\,K.\,J.\,B., W.\,B. and E.\,M.\,W. gratefully acknowledge financial support from the Deutsche Forschungsgemeinschaft
(DFG, German Research Foundation) through Project-ID 425217212 - SFB 1432. J.\,S.\,O. and E.\,M.\,W. further acknowledge funding from the European Union’s Horizon 2020 Research and Innovation Programme under Grant
Agreement No 732894 (FET Proactive HOT), and the German Federal Ministry of Education and Research
(contract no. 13N14777) within the European QuantERA cofund project QuaSeRT.
M.\,I.\,D. acknowledges support from the National Science Foundation, Grants No. DMR-1806473 and CMMI 1661618. M.\,I.\,D. is  grateful for the warm hospitality at the University of Konstanz and at the Technical University of Munich.   

%
%
\newpage
\appendix 
\section{Supplemental Material: Theory}
\subsection{Hamiltonian dynamics of a resonantly driven mode}

It is convenient to describe the Hamiltonian dynamics of a resonantly driven mode in two steps. The first step is the transition to the action-angle variables\addDB{,} $I$ and $\theta$\addDB{,} of the mode in the absence of driving. We relate these variables to the coordinate $q$ and momentum $p$ of the mode in the standard way \cite{Landau2004a} as 
\[I=(2\pi)^{-1}\oint p\,dq ,\qquad \theta = \frac{\partial}{\partial I}\int^q p\,dq,\]
with $q$ and $p$ being periodic in $\theta$,
\[q(I, \theta+2\pi) = q(I,\theta), \quad p(I, \theta+2\pi) = p(I,\theta). \]
An isolated mode performs periodic vibrations with $I=$~const. The phase linearly accumulates in time, $\dot \theta = \omega(I)$. The vibration frequency is $\omega(I) = dE/dI$, where $E$ is the mode energy.

We note that the full Hamiltonian of the system includes the kinetic and potential energy of the isolated mode that we consider, but also a contribution from other degrees of freedom, such as other nanomechanical modes of the resonator, acoustic phonons, etc. The mode is coupled to these degrees of freedom. This leads to a renormalization of the parameters of the mode, in particular, of the dependence of its vibration frequency on $I$. The function $\omega(I)$ incorporates this renormalization along with the harmonic and anharmonic terms of the potential energy $U(q)$ of the isolated mode.

In the presence of a resonant drive the Hamiltonian of the system takes the form
\begin{align}
\label{eq:SM_Hamiltonian_lab}
&H=H_0 + H_F, \qquad H_0=\int_0^I dI'\omega(I'),\nonumber\\
&H_F=-F q(I,\theta) \cos\omega_d t,
\end{align}
and the equations of motion read
\[\dot I = -\partial_\theta H_F, \quad \dot\theta = \omega(I) + \partial_I H_F.\]  

We consider the case where the driving is relatively weak, as explained in the main text. This means that the anharmonic part of the energy of the driven mode remains small compared to the harmonic part. In terms of $I$, the latter condition typically  means that 
\begin{align}
\label{eq:I_anh}
I\ll I_\mathrm{anh}, \quad |\omega(I_\mathrm{anh}) - \omega_0|\sim \omega_0\equiv \omega(0).
\end{align}
Here  $I_\mathrm{anh}$ is the value of the action variable where the change of the vibration frequency becomes comparable to $\omega_0$. 

The inequality (\ref{eq:I_anh}) is essentially the condition on the strength of the drive. We consider what we call a weak to a moderately strong drive or equivalently, small to moderately large vibration amplitude. This implies that,  for the characteristic values of $I$, the frequency change $F|\partial_I q|$  is small compared to $\omega_0$ as is also the rate of the change of the action $F|\partial_\theta q|$. However, the drive does not have to be small compared to the appropriately scaled frequency detuning $|\omega_d - \omega(I)|\ll \omega_0$. The interrelation between $F|\partial_I q|$ and $|\omega_d - \omega(I)|$ is arbitrary for typical values of $I$. For such a drive the system remains far away from the region where the dynamics becomes chaotic. 

The condition on the drive strength becomes more explicit if we expand $q(I,\theta)$ in $I$ and $\omega_d - \omega_0$. The leading-order term in the expansion is  
\begin{align}
\label{eq:q_of_I}
q(I,\theta) = (2I/M\omega_d)^{1/2}\cos\theta.
\end{align}
Here $(2I/M\omega_d)^{1/2}$ is the vibration amplitude for a given $I$. We note that nonlinear vibrations have overtones, which are disregarded in Eq.~(\ref{eq:q_of_I}). However, in the range of $I$ we are interested in, the amplitudes of the overtones are small compared to the amplitude of the main tone.  

To the order of magnitude, the vibration amplitude as a function of the drive amplitude can be estimated as $(2I_\mathrm{res}/M\omega_d)^{1/2}$, where $I_\mathrm{res}$ is given by the standard condition of nonlinear resonance \cite{Arnold1989,Ott2002} 
\[M\omega_d I_\mathrm{res} [\omega_d - \omega(I_\mathrm{res})]^2\sim F^2.\]
The drive is weak to moderately strong and the amplitude is small to moderately large  provided 
\begin{align}
\label{eq:moderate_drive_condition}
I_\mathrm{res} \ll I_\mathrm{anh}.
\end{align}

Equation (\ref{eq:q_of_I}) is the first term in the expansion of $q(I,\theta)$ in $I/I_\mathrm{anh}\ll 1$ and $|\omega_d-\omega(I)|/\omega_d\ll 1$. The smallness of the latter parameters justifies keeping the leading order term in this expansion in the range (\ref{eq:moderate_drive_condition}). We note that, in the same approximation, $p\approx -(2 I M\omega_d)^{1/2}\sin\theta$.


\subsubsection{Transition to slow variables}
\label{subsec:slow_variables}

To analyze the resonant dynamics we have to go to the rotating frame. This is done using the transformation
\begin{align}
\label{eq:variables_rotating_frame}
&q=Q\cos\omega_d t + P\sin\omega_d t,\nonumber\\
&p = M\omega_d (-Q\sin \omega_d t + P\cos\omega_d t).
\end{align}
In the approximation (\ref{eq:q_of_I}) we have 
\begin{align}
\label{eq:action_Q_P}
Q^2 + P^2 = 2I/M\omega_d.
\end{align}
We note that, in the approximation (\ref{eq:q_of_I})
\begin{align*}
&Q = (2I/M\omega_d)^{1/2}\cos (\theta - \omega_dt), \\
&P=-(2I/M\omega_d)^{1/2} \sin(\theta - \omega_d t),
\end{align*}
which shows that $Q$ and $P$ are slow variables, since $\dot \theta \approx \omega(I)$ and $\omega(I)$ is close to $\omega_d$.

The transformation (\ref{eq:variables_rotating_frame}) would be canonical if we used $Q/(M\omega_d)^{1/2}, P/(M\omega_d)^{1/2}$ instead of $Q,P$, but for the comparison with the experiment below it is more convenient to have $Q$ and $P$ of the same dimension as the coordinate $q$. As indicated in the main text, the variables $Q$ and $P$ correspond to the in-phase and quadrature components of the mode displacement. They slowly vary in time, remaining unchanged on the time scale $1/\omega_d$.

A time-dependent change of variables requires an appropriate change of the Hamiltonian \cite{Landau2004a}. In the present case, in the approximation (\ref{eq:action_Q_P}) the Hamiltonian becomes 
\[H'\equiv H'(Q,P)= (H- \omega_d I)/M\omega_d.\]
In fact, we have changed here from the variables $I,\theta$, to the variables $Q,P$, with $I$ related to $Q,P$ by Eq.~(\ref{eq:action_Q_P}).

The next step is averaging over the fast oscillating terms in $H'$. First of all, we note that $I\propto Q^2+P^2$ is a slow variable, fast oscillating terms in $H_0 - \omega_dI$ are small. The major fast-oscillating terms in $H'$ come from $H_F$. We evaluate them by substituting into $H_F$ the expression (\ref{eq:variables_rotating_frame}) for $q(I,\theta)$. Clearly, $H_F$ has the term $-QF/2$  with no time-oscillating factors and two terms that contain the factors $\cos 2\omega_d t$ and $\sin 2\omega_d t$, respectively. Averaging over the time $2\pi/\omega_d$ allows us to eliminate these fast oscillating terms. As a result the time-averaged Hamiltonian $H'(Q,P)$ takes the form $g(Q,P)$ given in Eq.~(3) of the main text, which we reproduce here for completeness:
\begin{align}
\label{eq:H_rotating_frame_SM}
&g(Q,P)=(M\omega_d)^{-1}(H_0-\omega_d I - QF/2),\nonumber\\
&I= M\omega_d (Q^2 + P^2)/2.
\end{align}
In quantum terms the eigenvalues of $g$ give the Floquet eigenvalues of the driven mode \cite{Shirley1965,Zel'dovich1967a,Ritus1967,Sambe1973}.
%


\subsubsection{The Hamiltonian for a small effective Duffing parameter}
\label{susbsec:working_g}

In the above approximation, the response curve of a resonantly driven nonlinear oscillator is determined by the dependence of the frequency $\omega(I)$ on the action variable. If we expand $\omega(I)$ to the second order in $I$,
\[\omega(I) = \omega_0 + \alpha_1 I + \alpha_2 I^2,\]
we have from Eqs.~(\ref{eq:SM_Hamiltonian_lab}) and (\ref{eq:H_rotating_frame_SM})
\begin{align}
\label{eq:g_explicit}
g(Q,P) &= -\frac{1}{2} \delta \omega \left(Q^2+P^2\right) + \frac{M\omega_d}{8} \alpha_1 \left(Q^2+P^2\right)^2 \nonumber\\
&+ \frac{(M\omega_d)^2}{24} \alpha_2 \left(Q^2+P^2\right)^3 - \frac{F}{2M\omega_d}Q.
\end{align}
Here 
\begin{align}
\label{eq:detuning_defined}
\delta\omega = \omega_d - \omega_0
\end{align}
is the detuning of the drive frequency from the eigenfrequency of the mode.

The Hamiltonian dynamics of the mode in the rotating frame is described by the equations of motion 
\begin{align}
\label{eq:Hamiltonian_eom}
\dot Q = \partial_P g(Q,P),\quad \dot P = -\partial_Q g (Q,P).
\end{align}
In the parameter range of interest this dynamics is vibrations with a given $g$, with frequency $\nu(g)$ that depends on $g$.

As explained in the main text, the parameters $\alpha_{1,2}$ in Eq.~(\ref{eq:g_explicit}) depend on the nonlinearity of the potential of the mode $U(q)$ and also on the nonlinear coupling of the mode to other modes of the resonator, to the electron system \cite{Moskovtsev2017}, and to other degrees of freedom. Therefore\addDB{,} calculating them requires knowing multiple parameters, and in fact the full characterization of the whole system, which is complicated if at all possible. This is essentially an unphysical task, as the parameters $\alpha_{1,2}$ are the only parameters that describe the resonant mode dynamics.

For completeness, we provide the expressions for $\alpha_{1,2}$ for an isolated mode with a nonlinear potential. We have to keep in this potential the terms up to the sixth order in $q$ to find $\alpha_{1,2}$,  
\begin{align}
\label{eq:potential}
U(q) &= \frac{M}{2}\omega_0^2 q^2 + \frac{M}{3} \gamma_3 q^3 + \frac{M}{4}\gamma_4 q^4 \nonumber\\
&+ \frac{M}{5} \gamma_5 q^5 + \frac{M}{6} \gamma_6 q^6\,.
\end{align}
The term $\propto q^4$ is the nonlinear term kept in the Duffing model. In the main text we used $\gamma$ instead of $\gamma_4$ as the coefficient of this term. Obviously, the nonlinear part of the potential contains 4 parameters, whereas only 2 parameters, $\alpha_1$ and $\alpha_2$ are actually accessible to the experiment.

It is more convenient to calculate the frequency as a function of energy $E$,
\begin{align}
\label{eq:omega_of_E}
\tilde \omega(E)& \equiv \omega[I(E)] \approx \omega_0 + \alpha_1\frac{E}{\omega_0} 
+\frac{E^2}{\omega_0^2}\left(\alpha_2 - \frac{\alpha_1^2}{2\omega_0}\right),
\end{align}
and then to find $\alpha_{1,2}$ from the expansion of $\tilde\omega(E)$ in a series in $E$. This expansion can be obtained from the expression
\begin{align}
\label{eq:integral_for_omega}
\tilde{\omega}(E) &= \pi \Bigl/\int\limits_{q_\mathrm{min}(E)}^{q_\mathrm{max}(E)} \frac{d{q}}{\sqrt{2[E-U(q)]/M}},
\end{align}
where $q_{\max}(E)$ and $q_{\min}(E)$ [$q_{\max} >q_{\min}$] are the turning points of the classical trajectory with a given $E$, i.e., $U(q_{\max}) =U(q_{\min})=E$. 
Introducing 
	\begin{align*}
		q_\mathrm{c} &= (q_{\max}+q_{\min})/2\,, & L &= (q_{\max}-q_{\min})/2\,, \\
		q &= q_\mathrm{c} + K\,, & K &= L \sin \alpha\,,
	\end{align*}
	we can write
	\begin{align}
		E-U(q(K)) &= \left(L^2-K^2\right) \left[\mu_0 + \sum_{n=1}^{4} \mu_n K^n\right]\,.
	\end{align}
Here, we used that the left hand side is a polynomial of degree six in $K$ with two real roots at $K=\pm L$ such that we can put the factor $\left(L^2-K^2\right)$ up front while the remaining part is a polynomial of degree four with the energy dependent coefficients $\mu_i$. This allows writing
	\begin{align}
		\int\limits_{q_\mathrm{min}(E)}^{q_\mathrm{max}(E)} \frac{d{q}}{\sqrt{2[E-U(q)]/M}} =  \frac{L}{\sqrt{2\left[E-U(q_\mathrm{c})\right]/M}} \nonumber\\
		\cdot \int_{0}^{\pi} \frac{d{\alpha}}{\sqrt{1+\sum_{n=1}^{4} \frac{L^2 \mu_n}{E-U(q_\mathrm{c})} L^n \cos^n(\alpha)}}\,.
	\end{align}
	Expanding $q_\mathrm{max}$ and $q_\mathrm{min}$ in powers of $E$ the square root can also be expanded in powers of $E$ and the integrals can be solved.
\begin{align}
	\label{eq:coefficients_for_potential}
	\alpha_1 &= \frac{3}{4M\omega_0^2} \left(\gamma_4 - \frac{10\gamma_3^2}{9\omega_0^2}\right)\,, \quad
	\alpha_2 = \frac{5 \gamma_6 }{4 M^2\omega_0^3} - \frac{7 \gamma_3 \gamma_5 }{2 M^2 \omega_0^5} \nonumber\\
	&- \frac{51\gamma_4^2 }{64 M^2 \omega_0^5} + \frac{75 \gamma_3^2 \gamma_4 }{16 M^2 \omega_0^7} - \frac{235 \gamma_3^4}{144 M^2 \omega_0^9}\,.
\end{align}
The parameter $\alpha_1$ is proportional to the well-known expression for the Duffing parameter $\gamma_4$ renormalized due to the cubic nonlinearity of $U(q)$ \cite{Landau2004a}. It is this renormalization (along with other terms that contribute to the renormalization) that can make $\alpha_1$ small, leading to the resonant response curve significantly different from the conventional Duffing curve.


\subsection{Dissipative dynamics}
\label{sec:dissipative}

The dynamics of the mode in the rotating frame  in the presence of dissipation is described by the equations \cite{Dykman2019}  
\begin{align}
\label{eq:eom}
&\dot Q = \partial_P g +R_Q, \quad \dot P = -\partial_Q g + R_P, \nonumber\\
&R_Q=  -(\Gamma + F^2 \Gamma_d) Q +\frac{\eta_\mathrm{RIFF}}{4M}FP^2,\nonumber\\
&R_P= -(\Gamma + F^2 \Gamma_d) P -\frac{\eta_\mathrm{RIFF}}{4M}FQP,
\end{align}
Here we have taken into account the change of the friction coefficient $\Gamma$ due to the driving-induced decay and described by Eq.~(5) of the main text. The term $\propto \eta_\mathrm{RIFF}$ describes the resonantly induced friction force (RIFF). We note that, for the microscopic mechanism of the RIFF considered in  \cite{Dykman2019}, the parameter $\eta_\mathrm{RIFF}$ is negative, and we will assume in what follows that $\eta_\mathrm{RIFF}<0$.

The decay rates in Eq.~(\ref{eq:eom}) should be compared with the frequency of oscillations $\nu(g)$ of the mode in the absence of decay. We will assume that $\nu(g)$ is much larger than the decay rate. This is the case in the experiment, where the decay rate is extremely small. We note that $\nu(g)\ll \omega_0$, that is, the decay rate is small not only in the laboratory frame, but also in the rotating frame.  

For a small decay rate, the motion is oscillations with a given scaled quasienergy $g(Q,P)=g$, with $g$ slowly evolving due to the decay. From Eq.~(\ref{eq:eom}), this evolution is described by the equation
\begin{align}
\label{eq:g_dynamics}
&\overline{\dot g} = \overline{\partial_Q g R_Q + \partial_P g R_P} \nonumber\\
&=\frac{\nu(g)}{2\pi} \int_{\mathcal{S}(g)} dQ dP (\partial_Q R_Q + \partial_P R_P).
\end{align}
Here the overline implies averaging over the trajectory $g(Q,P)=g$ and $\mathcal{S}(g)$ is the area inside this trajectory. We note that the sign in the last line refers to the trajectory that corresponds to the large-amplitude branch of the response curve; for the low-amplitude branch the sign is opposite.

\subsubsection{Stable states and limit cycles}
\label{subsec:stability}

It follows from Eqs.~(\ref{eq:eom}) and (\ref{eq:g_dynamics}) that, in the absence of the RIFF, we have $\dot g<0$ on the large-amplitude branch. The stable state of the system $(Q_\mathrm{st},P_\mathrm{st}$) then is the state with the minimal $g(Q,P)$, i.e., $g(Q_\mathrm{st},P_\mathrm{st})=g_{\min}$. It  is given by the equation
\begin{align}
\label{eq:stable_state}
[\partial_Q g(Q,0)]_{Q=Q_\mathrm{st}} = 0,\quad P_\mathrm{st}=0. 
\end{align}
It corresponds to the  amplitude of forced vibrations $A = Q_\mathrm{st}$. In fact, the stable state is slightly shifted from the minimum of $g(Q,P)$ because of the dissipation. The shift is small where the dissipation rate is small compared to $\nu(g_{\min})$. 

The solution of Eq.~(\ref{eq:stable_state}) was used to find the parameters $\alpha_1$ and $\alpha_2$ from the dependence of $A$ on  both the frequency and the amplitude of the drive measured in the experiment, in a broad range of these parameters. We note that, in the regime of very small amplitudes where the linear decay rate has to be taken into account, the vibration amplitude $A_0$ is different from $Q_\mathrm{st}$. To the leading order in $\omega_0 - \omega_d$ the response is described by
	\begin{align}
		\label{eq:full_response}
		&\frac{F^2}{4 M^2 \omega_0^2} = \nonumber\\  &A_0^2 \left[\Gamma^2 + \left(-\delta \omega + M \alpha_1 \frac{A_0^2 \omega_0}{2} + M^2 \alpha_2 \frac{A_0^4 \omega_0^2}{4}\right)^2\right]\,.
	\end{align}
For the values of $A_0$ where the term $\Gamma^2$ in Eq.~(\ref{eq:full_response}) can be disregarded, we have  $A_0 = A = Q_\mathrm{st}$. For small damping this happens already for a comparatively weak drive amplitude $F$.

For $A = Q_\mathrm{st}$, the decay rate $\lambda$ near the stable state is
\begin{align}
\label{eq:decay_rate}
\lambda = \Gamma + F^2 \Gamma_d + (\eta_\mathrm{RIFF}/8M)FQ_\mathrm{st}.
\end{align}
This decay rate gives the halfwidth of the sideband peaks in the power spectrum of the driven mode, as seen in \cite{Huber2020}.
In the region where $F^2\Gamma_d$ can be disregarded,  the rate $\lambda$ decreases with the increasing drive amplitude $F$ for $\eta_\mathrm{RIFF}<0$, since $Q_\mathrm{st}$ as well as the factor $F$ itself in the last term increase with $F$. Overall the decrease of $\lambda$ with the increasing $F$ is superlinear. This is in agreement with the decrease of the spectral linewidth in Fig.~2~(b) of the main text.

The value of the drive parameters where $\lambda=0$ corresponds to the Hopf bifurcation \cite{Guckenheimer1997}. As the drive amplitude or frequency further increase, the state where  $g=g_{\min}$ becomes unstable. The stable state is a limit cycle, which for a small decay rate, is given by the equation $g(Q,P) =g$ with $g$  given by the condition $\overline{\dot g}=0$, or 
\begin{align}
\label{eq:stable_limit_cycle}
&\Gamma + F^2 \Gamma_d = -\frac{\eta_\mathrm{RIFF} F}{8M}\,\overline{Q}(g),
\nonumber\\
&\overline{Q}(g) =\frac{1}{\mathcal{S}(g)}\int_{\mathcal{S}(g)} QdQ \,dP, \quad  
\mathcal{S}(g)\equiv\int_{\mathcal{S}(g)} dQ \,dP.
\end{align}
That is, the limit cycle is very close to the Hamiltonian trajectory (\ref{eq:Hamiltonian_eom}). This trajectory depends on the drive parameters.

The mean coordinate in the rotating frame $\overline{Q}(g)$  decreases with the increasing $g$. Its value is maximal for $g=g_{\min}$. The system is unstable if the right-hand side of Eq.~(\ref{eq:stable_limit_cycle}) exceeds the left-hand side for $g=g_{\min}$. But as $g$ increases $\overline{Q}(g)$ falls off \cite{Dykman2019} and ultimately the condition (\ref{eq:stable_limit_cycle}) is met.  

The trajectories $g(Q,P)=$~const with the value of $g$ given by Eq.~(\ref{eq:stable_limit_cycle}) are plotted in Figs.~3~(b) and 5~(a) of the main text. An important feature of these trajectories is that they are profoundly non-elliptical. Therefore\addDB{,} they have multiple Fourier components. It is this feature that leads to the onset of the frequency comb in the laboratory frame.


\subsection{The frequency comb in the power spectrum}

We now discuss the power spectrum of the mode where, in the rotating frame, it vibrates with a given value of $g(Q,P)=g$, i.e., with a given quasienergy. The spectral density of fluctuations of the displacement $q(t)$ of the resonantly driven mode near the driving frequency $\omega_d$ has the form 
\begin{align}
\label{eq:spectrum_general}
S(\omega) &= \frac{1}{2t_l}\left\vert\int_{-t_l}^{t_l}dt\,q(t)e^{i\omega t}\right\vert^2\nonumber\\
&\approx \frac{1}{8t_l}\left\vert\int_{-t_l}^{t_l}dt\,[Q(t)+iP(t)]e^{i(\omega-\omega_d) t}\right\vert^2,
\end{align}
where it is implied that $t_l\to \infty$. We have assumed that $|\omega-\omega_d|\ll \omega_d$ and expressed $q(t)$ in terms of the slowly varying in time quadratures $Q(t), P(t)$ using Eq.~(\ref{eq:variables_rotating_frame}).

Using the Hamiltonian equations of motion (\ref{eq:Hamiltonian_eom}) for $Q(t), P(t)$,  we can write
\begin{align}
\label{eq:Fourier_introduced}
[Q(t) +iP(t)]_g = \sum _mz_m (g) e^{im\nu(g) t}. 
\end{align}
Here, $[\cdot]_g$ indicates that the value is evaluated for a given $g(Q,P)$ and  $\nu(g)$ is the oscillation frequency in the rotating frame. The Fourier components $z_m$ are also determined by the value of $g$. 
For a resonantly driven Duffing oscillator, $\alpha_2=0$, they were calculated and used earlier \cite{Dykman1988a,Dykman2005,Guo2013} taking into account  that, in this case, the trajectories (\ref{eq:Hamiltonian_eom}) are expressed in terms of the  Jacobi elliptic functions. The parameters $z_m$ were also given in Ref.~\cite{Dykman2019}; however, the expressions of the latter paper for $m<0$ need to be corrected to account for the proper parallelogram of periods of the relevant functions.

From Eqs.~(\ref{eq:spectrum_general}) and (\ref{eq:Fourier_introduced}), the  power spectrum $S(\omega)=S_g(\omega)$ of the driven oscillator for a given $g(Q,P)$ is  
\begin{align}
\label{eq:spectrum_Fourier}
S_g(\omega) = \frac{\pi}{2}\sum_m|z_m(g)|^2\delta[\omega-\omega_d+ m\nu(g)].
\end{align}
The spectrum (\ref{eq:spectrum_Fourier}) is a frequency comb. It consists of a set of equidistant peaks separated by $\nu(g)$. The intensity (area) of the peaks is given by the Fourier components $z_m(g)$; note that, generally, $z_{-m}(g)\neq z_m^*(g)$.

For $\alpha_2\neq 0$ the Fourier components  $z_m(g)$ were found from the numerical solution of Eq.~(\ref{eq:Hamiltonian_eom}). The results for $|z_m(g)|$ are presented in Figs.~3~(c) and 5~(b) of the main text.

We note that below the threshold of instability the mode performs fluctuation-induced vibrations about $Q_\mathrm{st}, P_\mathrm{st}$. They lead to peaks in the power spectrum at frequencies $\omega_d\pm \nu(g_{\min})$. Slightly above the threshold the stable value of $g$ as given by Eq.~(\ref{eq:stable_limit_cycle}) is close to $g_{\min}$. The vibrations in the rotating frame are nearly sinusoidal in this range and their frequency is close to $\nu(g_{\min})$. Therefore, the major spectral manifestation of going through the threshold is the narrowing of the spectral lines, as indeed seen in the experiment.

\section{Supplemental Material: Calibration}
In the experiment the drive amplitude $F$ is not directly accessible. Instead, the RF input voltage $V_\mathrm{in}$ that determines the amplitude of the AC voltage is controlled. Similarly, the position of the mode $q$ and its vibration amplitude at the drive frequency, $A$, are not directly measured, but the output signal at the drive frequency, $V_\mathrm{out}$, is measured. We assume that both of these voltage signals relate to the physical quantities in a linear way such that
\begin{align}
	A &= a V_\mathrm{out}\,, & F/M &= b V_\mathrm{in}, \nonumber
\end{align}
with $a$ and $b$ being calibration constants.

For weak drives the force dependent friction terms, Eqs.~(4) and (5) in the main text, can be neglected and the response of the system is described by Eq.~(\ref{eq:full_response}). Expressing in this equation $F$ and $A$ in terms of $V_\mathrm{in}$ and $V_\mathrm{out}$, we find that the needed model parameters are the conversion factor 
\[c = b^2/4\omega_0^2 \Gamma^2 a^2,\]
the linewidth $2 \Gamma /2\pi$, the frequency $\omega_0$, as well as the nonlinearity parameters $M\alpha_1 a^2$ and $M^2 \alpha_2 a^4$. 

The parameters can be determined from fits in a systematic way, making use of the fact that, for small drive powers and therefore small amplitudes, the contribution from the term with $\alpha_2$, and for very small drive powers also the contribution from $\alpha_1$, can be neglected. Assuming a constant noise floor of $1.5\cdot 10^{-6}$\,mV, we fit the different regimes. We start at a drive power of  $-56$\,dBm with the Lorentzian response, proceed to a drive power of $-37$\,dBm with the Duffing fit, where we take the contribution from $\alpha_1$ into account, and finally fit the response at $-24$\,dBm with the full Eq.~(\ref{eq:full_response}) to determine $\alpha_2$.

Except for the plots in Figs.~3~(b) and ~5~(a) of the main text, all data and theoretical calculations belong to the same set of calibration parameters. For the data of the trajectory measurements, namely the data shown in Fig.~3~(b) and Fig.~5~(a), the conversion factor $a$ is different. Using a second set of response measurements in the different regimes we reconstructed the ratio between the two $a$ factors by comparing the fit values for $\tilde{\alpha}_1$ between the two sets of response measurements. We used their ratio $r$ to scale the data of Fig.~3~(b) and Fig.~5~(a).

The phenomenological parameters of the dissipation mechanisms are fixed in a separate step. Seeking for the best description of the amplitude of the vibrations at the drive frequency as a function of the drive power shown in Fig.~2~(c), we set $\Gamma_d$ to some value and fix $\eta_\mathrm{RIFF}$ by demanding that the Hopf bifurcation occurs at the threshold power $\mathcal{P}_\mathrm{th}$, meaning that the parameter $\lambda$ given by Eq~(\ref{eq:decay_rate}) goes to zero at $\mathcal{P}_\mathrm{th}$.

The values we find are
\begin{align*}
	c &= 0.00528 & \omega_0\,, &= 4.10 \cdot 10^7\,\mathrm{/s}\,,\\
	2\Gamma/2\pi &= 21.1 \mathrm{/s}\,, & M a^2\alpha_1 &= 45.2\,/\mathrm{V}^2\,, \\
	M^2 a^4 \alpha_2 &= 0.400\,\mathrm{s}/\mathrm{V}^4\,, & a^2 \eta_\mathrm{RIFF} &= 2.81 \cdot 10^{-3}\mathrm{s}/\mathrm{V}^2\,,\\
	M^2 a^2 \Gamma_d &= 4.5 \cdot 10^{-16} \mathrm{s}^3/\mathrm{V}^2\,, & r &= 1.21\,.
\end{align*}

\section{Supplemental Material: Additional experimental observations}
\subsection{Mechanical eigenmodes}

Several measurements have been performed to verify that only one mechanical mode, the fundamental out-of-plane (OOP) mode of the $55\,\si{\micro \meter}$ long nanostring under investigation ($f_{0}=6.529$\,MHz) is involved in the frequency comb formation. 

\begin{figure}[t]
	\centering
	\includegraphics[width=0.9\linewidth]{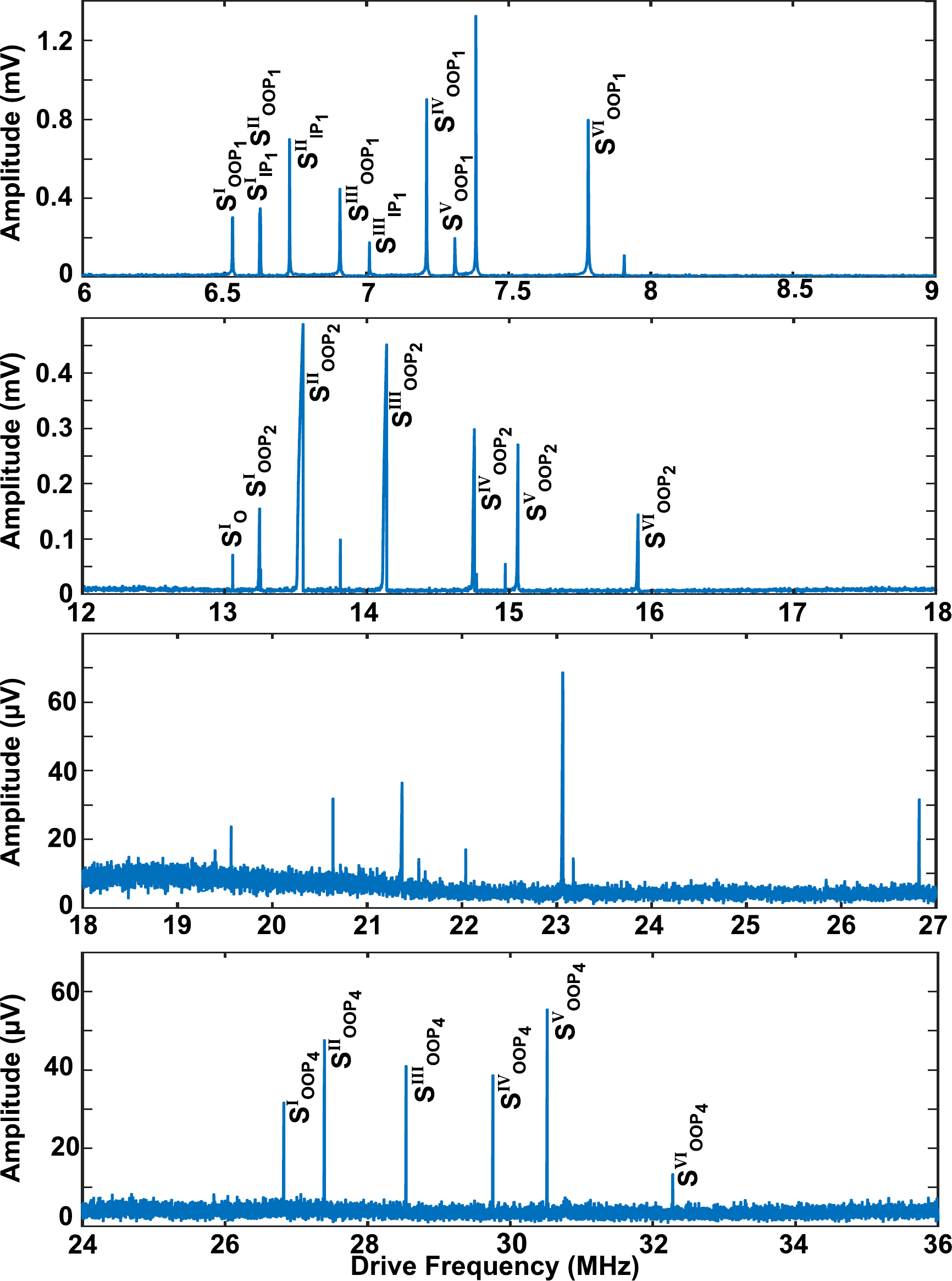}
	\caption{Mode spectra of the investigated sample hosting six working nanostring resonators. The applied drive powers are $-20$\,dBm, $-10$\,dBm and $-4$\,dBm, from top to bottom;  the lowest two panels were both measured for a drive power of $-4$\,dBm. The lowest-frequency mode at approx. $6.5$\,MHz is the OOP mode of the longest string discussed in this work. Identified modes are labelled by the string number (S$^I$ to S$^{VI}$, from the longest to the shortest working nanostring), the mode polarization (OOP or IP for out-of-plane and in-plane modes), and the harmonic ($1$ to $4$ from the fundamental up to the fourth harmonic). The label S$^I_O$ refers to the second overtone at exactly twice the eigenfrequency of the OOP mode at $6.5$\,MHz. It does not correspond to an eigenmode of the system. A few spurious modes that could not be identified unanbiguously remain unlabelled.}
	\label{fig:5modes}
\end{figure}

\begin{figure}[h]
	\centering
	\label{fig:modessingledrive}
	\includegraphics[width=0.9\linewidth]{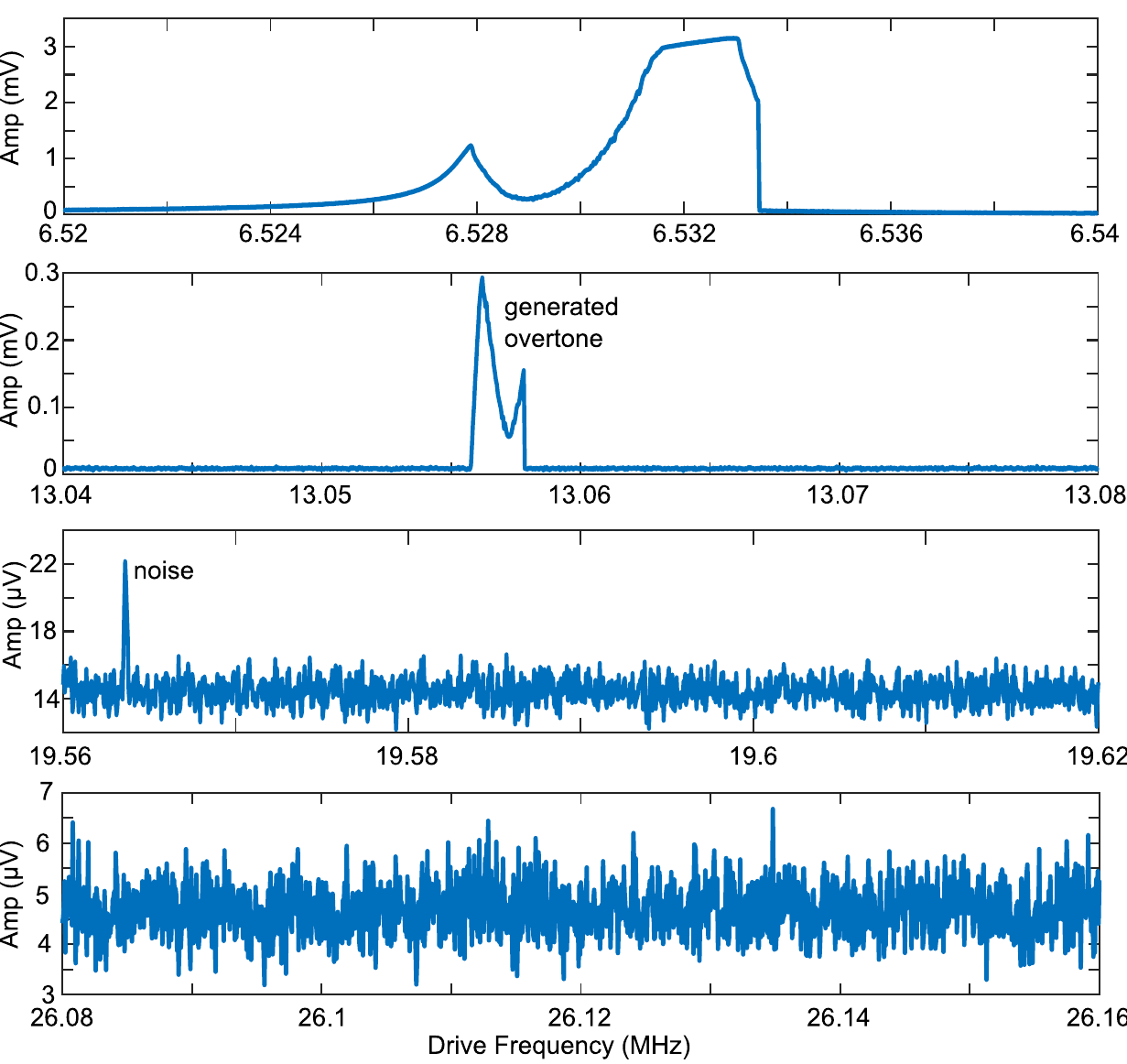}
	\caption{Frequency response measurement in the RIFF regime (drive power $0$\,dBm), sweeping the drive frequency through the eigenfrequency of the OOP mode. The four lines show the response of the sample at the drive frequency, as well as at twice, three and four times its value.}
	\label{fig:modes_sweep}
\end{figure}

Generally, a DC voltage applied to the dielectric control electrodes tunes the eigenfrequencies and can induce an appreciable coupling between the out-of-plane and in-plane mode of the nanostring~\cite{Rieger2012,Faust2012a}. However, as already discussed in the main text, all measurements in this work have been done at a constant DC voltage of $5\,\si{\volt}$ where all modes are tuned sufficiently far from resonance so that the coupling is effectively weak and the out-of-plane and in-plane mode can be considered as independent eigenmodes of the system. 

Besides higher harmonic eigenmodes, also the modes of additional nanostrings on the sample need to be considered to obtain a complete picture. The sample hosts a series of $12$ nanostring resonators with lengths ranging from $33\,\si{\micro \meter}$ to $55\,\si{\micro \meter}$, which are shunted between the same pair of control electrodes. Out of the twelve resonators, six are working, including the longest one which is the nanostring we are focussing on in this work. As it is the longest nanostring on the sample its OOP mode is the lowest frequency eigenmode found. Note that the fundamental in-plane eigenmode occurs at a slightly higher eigenfrequency as a result of the nanostring's width slightly exceeding its thickness. 
The full mode spectrum of the sample is shown in Figure \ref{fig:5modes}. The drive frequency was swept up to a frequency of $36$\,MHz while recording the response of the forced vibrations at the drive frequency. The measurement clearly demonstrates that while there are many mechanical modes on the sample, there is no $1:n$ internal resonance between the OOP mode under investigation and any other of the modes.

Figure \ref{fig:modes_sweep} shows a frequency response measurement where we sweep the drive frequency around the eigenfrequency of the OOP mode under investigation in the regime of self-sustained vibrations in the rotating frame, along with the response of the system at twice, three and four times the drive frequency. The only signatures found at higher frequencies are the second overtone at exactly twice the drive frequency and a spurious noise peak 
at  $\approx 19.56$\,MHz which is not associated with the drive. This measurement confirms and extends the conclusion drawn from Fig.~1(d) of the main text that no additional modes are nonlinearly excited, even when driving slightly off resonance.

%
%

%

%
\end{document}